# Bridging adaptive management and reinforcement learning for more robust decisions

Melissa Chapman[1], Lily Xu[2], Marcus Lapeyrolerie[1], Carl Boettiger[1]

[1] Dept. of Environmental Science, Policy, and Management, University of California, Berkeley, Berkeley, CA, USA
[2] School of Engineering and Applied Sciences, Harvard University, Cambridge, MA, USA

**Abstract:** From out-competing grandmasters in chess to informing high-stakes healthcare decisions, emerging methods from artificial intelligence are increasingly capable of making complex and strategic decisions in diverse, high-dimensional, and uncertain situations. But can these methods help us devise robust strategies for managing environmental systems under great uncertainty? Here we explore how reinforcement learning, a subfield of artificial intelligence, approaches decision problems through a lens similar to adaptive environmental management: learning through experience to gradually improve decisions with updated knowledge. We review where reinforcement learning (RL) holds promise for improving evidence-informed adaptive management decisions even when classical optimization methods are intractable. For example, model-free deep RL might help identify quantitative decision strategies even when models are nonidentifiable. Finally, we discuss technical and social issues that arise when applying reinforcement learning to adaptive management problems in the environmental domain. Our synthesis suggests that environmental management and computer science can learn from one another about the practices, promises, and perils of experience-based decision-making.

**Keywords:** adaptive management; artificial intelligence; reinforcement learning; deep learning; biodiversity; decision making

# Introduction

Given the urgency of environmental crises and the impending risk of crossing planetary tipping points, developing management strategies in the face of uncertainty is increasingly critical. Decision making under uncertainty, central to most contemporary environmental policy and practice, is referenced in contexts that span scales and systems, ranging from multilateral initiatives to halt biodiversity loss (1) and mitigate climate change (2) to local measures for habitat protection (3) and water allocation (4). But uncertainty - and the possibility of triggering regime shifts - persist as challenges to effective conservation decision-making (5), motivating conservation science, policy, and practice to focus not only on making decisions given uncertainty but iteratively *reducing* uncertainty through adaptive decision-making.

Since the coining of the term *adaptive management* in the 1970's (6), attempts to apply the paradigm of "learning while doing" have proliferated across the environmental domain (Box 2) (7,8). But the theoretic underpinnings of adaptive management - reducing uncertainty around a discrete set of autonomous models (systems that do not explicitly depend on independent variables) or model parameter values over time to take actions that maximize a notion of expected utility - have proven to be difficult in application (9,10), buckling under the problem complexity and sociopolitical nuance within which real environmental decisions are made (11,12). Notably, the decision-theoretic methods commonly used to solve adaptive management problems (e.g., Bayesian model updating and dynamic programming) generally assume that uncertainties are not only known but can be precisely quantified in probabilities (13). This, in turn, hinges on the assumption that models of environmental systems are identifiable and autonomous, which is often not true, particularly in the case of systems with tipping points or in the context of rapid environmental changes due to anthropogenic pressures (14).

In light of the limitations of decision-theoretic methods, multiple alternative approaches have emerged to inform the management of complex systems under uncertainty (e.g., scenario planning and resilience thinking (5)) For example, statistical early warning signals of tipping points (e.g., critical slowing down) allow us to punt on the issue of model misspecification. Notably, early warning signals, scenario planning, and other "resilience thinking" approaches all shift away from quantifying a decision policy (e.g., a model which suggests a fishing quota of X metric tons) and towards classifying (e.g., "the system is / isn't approaching a tipping point", or "scenario B is preferable to scenario C"). Hereafter, we will call these "classification" approaches. Despite long-standing calls to better integrate decision theory and classification approaches (13), classification has continued to retreat from the decision problem all together, focusing instead on advanced computational tools - like deep learning - to better predict complex systems dynamics while avoiding the need to explicitly consider actions or the expected utility of a given strategy. For example, (15) uses machine learning to classify critical transitions into four possible classes but does not suggest action strategies given a belief state in the system dynamics.

At first glance, focusing on system classification is compelling: if we can identify a proximate tipping point or predict a threshold response in a system - regardless of the exact underlying model (e.g.,(15)) - surely a manager can leverage that knowledge to compose a good management strategy? But is washing our hands of devising quantitative decision strategies for complex systems really a good idea? Heuristic approaches to decision-making notably fail to design effective sequential decision strategies (in contrast to iterative or static decision strategies) compared to formal approaches (16). This leads us to ask if the limitations of a heuristic decision are better than those faced by traditional model-based decision theory (17).

If we can predict thresholds more successfully without models, can we also derive quantitative decision strategies for those systems more effectively without models (and the assumptions of model-based decision-theoretic approaches)? Within the field of computer science, a rapidly emerging class of machine learning algorithms has proven remarkably effective at making complex sequential decisions without first learning a model of the system (18). Interestingly, these model-free reinforcement learning (RL) algorithms often mimic what any good manager does in the same situation: forgoing learning over an explicit set of process-based models and relying instead on *knowledge from successes and failures experienced by repeatedly making decisions*. But unlike a human manager, RL algorithms can process near real-time and high-dimensional data as well as learn strategies by interacting with a wide diversity of simulations and scenarios (19) that would be infeasible for a human to process. Moreover, RL algorithms show promise to more effectively manage systems with unknown tipping points than fixed "rule-of-thumb" strategies (20).

In this paper, we first highlight the limitations and assumptions of decision-theoretic approaches that have driven a wedge between the theory and practice of adaptive environmental management. We then explore examples of successful reinforcement learning applications, highlighting which specific RL methods and concepts might provide a promising path forward to overcome some limitations of model-based adaptive management (Table 1). Despite the promise of these emerging methods, the problem of making strategic decisions in nonidentifiable and/or nonautonomous (systems subjected to external inputs) systems is not easily solved by even the best algorithms. We simultaneously suggest that RL research might benefit from the insights gleaned from trying to tackle pressing environmental problems (Table 1). Finally, we emphasize that while RL approaches may, at last, allow algorithms to help human decision-makers better grapple with real-world complexities, such a transition will raise new challenges for equity, governance, and accountability.

## Adaptive management: translating theory to practice

Given the limitations of model-based approaches to adaptive management (Box 1), the practice of adaptive management - and its integration into policy - has deviated from formal definitions of decision theory (Figure 1). Rather, implemented adaptive management strategies often take "rule-of-thumb" approaches, substituting human judgment and the experience of decision-makers in place of computationally intensive models (7,21). While more loosely defined and human-informed adaptive management processes overcome some of the technical shortcomings of model-based adaptive management (e.g., by allowing for consideration of greater system complexity and the social context of decisions), these approaches also face several limitations. Beyond the myriad of potential non-technical risks, such as goal slippage, manager turnover, and sensitivity to political asymmetries and institutional change (22), "rules of thumb" face technical constraints. For example, the exponentially increasing amount of monitoring data from automated sensors (23) quickly becomes impractical for humans to integrate into a heuristic decision process.

Thorough reviews of the application (7,8), theory (24), and legal implications (25) of adaptive management already exist in the literature. In Figure 1A-B, we summarize the adaptive management paradigm. It should be noted that while adaptive management is not appropriate in all settings, the benefits of leveraging adaptive over nonadaptive approaches are particularly apparent in contexts where there is deep uncertainty about system dynamics—a setting in which RL might hold the most promise. In the remainder of this section, we illustrate, through examples, the key gaps between adaptive management as it is expressed in theory and the requirements for its effective implementation in practice. While decision-makers deviations from scientific and/or algorithmic recommendations and definitions of adaptive management (Figure 1B) are often viewed with skepticism, we explore how this deviation frequently reflects the inability of models integrated into decision-theoretic frameworks to sufficiently account for real-world complexity and challenges (Box 1).

Take, for example, the Tallapoosa River, where an upstream hydropower dam had uncertain effects on the integrity of the river's species-rich ecological community, prompting a decade-long adaptive management effort (9,26) (Box 1A). Monitoring how multiple ecological indicators and stakeholder preferences respond to flow alterations was used to iteratively update a model (a state-space transition model) of the system, which was integrated into a decision-theoretic framework to optimize flow regimes (26). While the river monitoring effort held promise to instruct more responsive and informed decisions, the high-dimensional system (which included multiple species and ecosystem indicators, recreational interests, dam revenue, seasonal temperature) and nearly unlimited potential flow regime strategies were reduced to three system

"states" and four potential flow regime "actions" over which the formal adaptive management problem learned and optimized (26). Would a more realistic representation of the systems, at least mirroring the scope of monitoring data collected and a more extensive set of possible actions, have allowed the solutions to capture nuanced and temporal tradeoffs between human use and ecological demands in this system?

Unfortunately, what adaptive management models add in social and ecological complexity they often lose in the number of models and parameters over which they learn due to computational constraints. For example, in the case of adaptively managing horseshoe crab harvests in the Delaware Bay (Box 1C), proposed frameworks included multi-species population models and a broader set of potential actions but consider only two competing models over which to reduce uncertainty and derive decision strategies (27). Beyond the challenges of capturing complexity and uncertainty of system states and actions, model-based adaptive management frameworks face limitations to inform decision-making in the context of global environmental change (24). For example, in the case of adaptively setting quotas for waterfowl harvest in the United States (Box 1B), the increasingly nonautonomous ("nonstationary") nature of the system (due to climate change not being included as part of the population model) is beginning to limit the applicability of the model-based decision theoretic methodology to devising decision strategies (28). Of course, climate dynamics could be integrated into the models. However, this further complicates both the model and the computational requirements to solve for optimal decision strategies using dynamic programming, as well as introduces additional uncertainty.

In each of the cases in Box 1, human decision-makers wrestle with a growing number of complexities, competing needs, and pressures left unanswered by the existing decision theoretic adaptive management paradigms. These mismatches limit the utility of existing model-based optimization approaches to the decision-making process and risk unintended consequences of relying on best-fit models in decision processes. Importantly, proposals to leverage adaptive management for higher-dimensional environmental problems, such as climate mitigation (29) and protected area design and management (30,31), are likely only to widen the gap between model-based theory and the realities of management decisions in practice. While classification approaches enable skirting the decision problem all together, less constrained approximate methods for finding optimal strategies have also been suggested for dealing with high-dimensional natural resource management problems for decades (32). In the following sections, we explore how emerging methods from deep reinforcement learning might allow us to leverage the benefits of both heuristic and computational approaches to adaptive management, improving our capacity to manage systems under uncertainty (Table 1, Figure 1C).

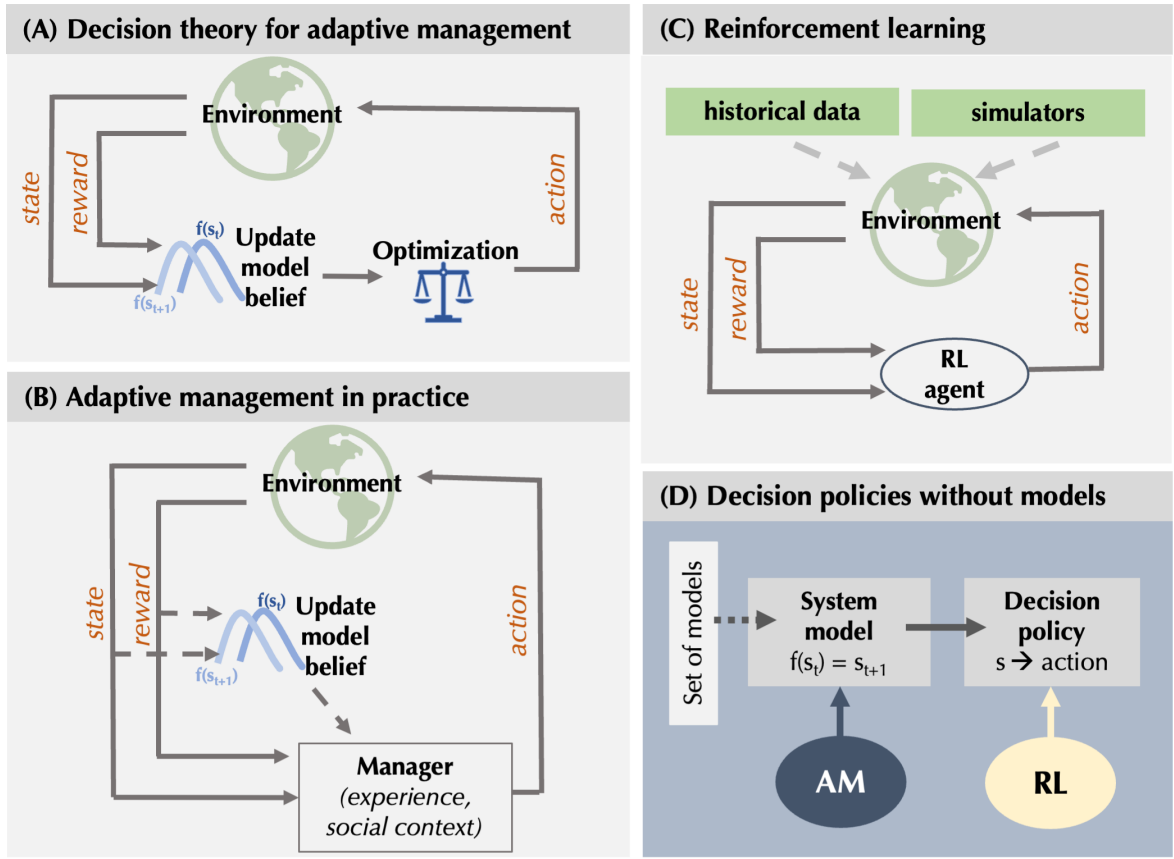

**Figure 1: (A)** Decision theoretic approaches to adaptive management - what we refer to as "model-based" adaptive management - formulate problems as Markov decision processes (MDPs) with unknown state transition probabilities (for a helpful overview of MDPs for ecology, see (16,33)). At each time step *t*, the agent takes an action *a*, on an environment, and the environment transitions from the current state *s*, to the next state $s_{t+1}$ and provides the agent with a reward *r*. After taking an action, the agent observes the system state and rewards to update its belief of the underlying dynamics of that system to inform the decision strategy ("policy") which selects the next action. **(B)** While theory provides a formalized way to "learn" about systems while acting on them, in practice, adaptive management decisions are usually made by humans who may or may not use the output of models to support decisions but often rely on experience with the complex systems and sociopolitical context not captured by stylized system models. **(C)** Like classical model-based optimization approaches, the task for a reinforcement agent is to learn a "policy" that maps the state of the system to an action the agent should take to maximize the expected sum of future rewards (29). This can be done by interacting with historical data or simulators. However, the process by which RL learns optimal policies is fundamentally different from (A). **(D)** Unlike the classical methods used in model-based adaptive management, reinforcement learning allows action strategies to be developed without learning over a predetermined set of potential models.

**Box 1:** Examples of model-based adaptive management and their key challenges.

| (A) Adaptive Flow Regimes: Tallapoosa River | |
|---|---|
| For nearly a decade, the United States Geological Survey (USGS) implemented adaptive management on the Tallapoosa River to determine optimal flows under multiple competing stakeholder objectives. The iterative decision model included annual monitoring of ecosystem indicators to vary flow regimes optimally. Despite the adaptive decision framework, the best management strategies to provide both adequate hydrologic and thermal habitats while sufficing recreational values remain a central controversy in the system (26). | **Objective:** ensure the conservation of at-risk species and meet ecosystem service objectives<br>**Learning:** passive learning of ecosystem dynamics and the valuation of ecosystem services<br>**States:** discrete set of ecosystem indicators, ecosystem values<br>**Actions:** four flow allocation strategies<br>**Key challenges faced:** multiple competing objectives, noisy observations, delayed feedbacks, stochasticity, high dimensionality |
| **(B) Adaptive Harvest Management: Waterfowl** | |
| Since 1995, the United States Fish and Wildlife Service has used an adaptive management framework to regulate duck harvests. Harvest quotas rely on an iterative cycle of monitoring, assessment, and decision-making. Based on monitoring data, managers continually refine models of the relationships between hunting regulation, harvests, and waterfowl abundance. While significant updates to the model weights have occurred over the past 20 years, non-stationarity due to global change challenges the current methodology (28). | **Objectives:** sustainable harvest of waterfowl<br>**Learning:** passive learning of population dynamics in response to harvest regimes<br>**States:** waterfowl abundance<br>**Actions:** annual harvest quotas<br>**Key challenges faced:** non-stationarity, delayed feedbacks, stochasticity, multiple objectives |
| **(C) Adaptive Harvest Management: Red Knots (*Calidris canutus rufas*) and Horseshoe Crabs (*Limulidae polyphemus*)** | |
| Following increased harvest of horseshoe crabs in the Delaware Bay during the 1990s, migratory shorebird populations declined steeply. Recognizing this decline, the fisheries commission began regulating the horseshoe crab harvest. Proposed adaptive management frameworks focus on two competing models of red knot population dynamics and horseshoe crab harvests, seeking to iteratively improve harvest policies for both objectives (27). | **Objective:** ensure the conservation of at-risk species while also meeting harvest objectives<br>**Learning:** passive learning of multispecies dynamics<br>**States:** red knot abundance and fecundity, horseshoe crab abundance<br>**Actions:** harvest quota for horseshoe crabs<br>**Key challenges faced:** model set limitations, dimensionality, delayed feedbacks, multiple objectives |

# Reinforcement learning as model-free adaptive management

Reinforcement learning has proven better than humans at making strategic, adaptive, and complex decisions across a diversity of problems and domains. While RL's most cited feats are in the context of games (e.g., chess) (34) and robotic tasks (35), RL algorithms are increasingly used to solve planning problems across a variety of noisy and uncertain real-world settings, from healthcare (36) and energy systems (37) to biological systems (38) and economic policy (39).

The problem setup for reinforcement learning closely mirrors model-based adaptive management (Figure 1). Like classical model-based optimization approaches (16), an RL agent aims to learn a "policy" (decision strategy) that maps the state of the system to the best action to take to maximize the expected sum of future rewards (20). However, the process by which RL learns optimal policies can be fundamentally different. Unlike classical dynamic programming methods leveraged in adaptive management examples from Box 1, which require specifying state transition matrices (state-space models of the system over which to learn) (16) (Box 1, Figure 1A, Figure 1D), *model-free* reinforcement learning allows action strategies to be developed without a predetermined model (Figure 1C), bypassing the need to iteratively learn a 'best model' of system dynamics altogether (Figure 1D). Importantly, RL learns action strategies through experience, which can include simulated experience, experience derived from historical data, and/or real-world experience.

First, RL can learn from **simulated experience**. A notable example is Atari, where researchers achieved human-level performance across dozens of Atari video games by training RL agents over millions of timesteps (decisions) (19) (Box 3A). Instead of fine-tuning a new model for each of the game, a single RL agent using the same neural network architecture and hyperparameters was applied to 49 different Atari games, reaching performance comparable to a professional human game player across a majority of the games. Without the goal of learning a single system transition matrix, the RL agent learns generic concepts that allow for strategic decision-making in a diversity of settings. Beyond gameplay, simulators can also model physical environments such as atmospheric wind conditions, as used to build a simulator to train an RL agent to navigate high-pressure balloons in the stratosphere (Box 2B). In the context of adaptive management, simulators could be designed to describe water flow, species population dynamics, or a changing climate. Leveraging simulated experience to learn policies for adaptive management problems could provide a means to integrate more complex system dynamics and action sets, such as in nonautonomous environments.

While RL allows us to relax assumptions required by most model-based adaptive management frameworks (e.g., MDPs), managing nonautonomous systems remains a hard problem for both computer and environmental science. We do not suggest that RL will readily overcome this challenge, but rather that in complex socio-ecological settings (e.g., the Tallapoosa River, Box 1A), RL might outperform both human heuristics and model-based adaptive management methods through exploring a set of complex simulated environments millions of times, accruing orders of magnitude more feedback than a system that could only interact with the real environment (40).

Notably, simulated experience is not the only way RL can learn decision strategies. Deep RL has been shown to learn effective policies from **historical data** in the absence of a simulator by stitching together trajectories of system observations ("offline reinforcement learning") (41,42). For example, in hospitals—which often have decades of historical records about patient status, treatment, and outcomes—this offline approach was taken to learn individualized treatments for sepsis patients (Box 2C). Of course, the extensive data available in some societal domains like healthcare is less common in environment systems. In some environmental challenges, such as water and air quality monitoring, sensors are already constantly taking samples potentially allowing for the tracking of the impact of specific actions. While this level of monitoring data is not available for many other ecological management problems, improved sensing of everything from vegetation dynamics to species occurrence (43) are trending the field in that direction and likely making RL-based management more feasible.

A key challenge in the environmental domain will be *off-policy evaluation*, to estimate the performance of policies that were never observed in the offline data (44). A large body of techniques for off-policy evaluation have been developed for RL for observational healthcare (45). Additional observations collected through real-world ("online") experience may then be used to improve the policy further or "adaptively" update policies while taking actions.

Emerging methods in RL have the potential to address more than just the issue of dimensionality and nonautonomous dynamics in natural resource management problems. From making effective decisions in environments with sparse rewards to addressing systems with multiple competing objectives, in Table 1, we map adaptive management challenges to RL methods that might help address them.

**Box 2:** Examples of reinforcement learning applications and the key challenges they address.

| **(A) Playing strategic games:** Atari ||
|---|---|
| Dealing with high-dimensional inputs to make effective decisions across different tasks and situations remains a key challenge for RL. Atari consists of 49 distinct video games with visual inputs and has become a go-to benchmark for developing RL algorithms. Using deep Q-learning, a single RL agent was trained across dozens of games to outcompete human players (19). | **Key challenges addressed:** high-dimensional input, diverse tasks, long horizon<br>**Objective:** maximize score across a set of Atari games<br>**State:** 84 × 84 × 4 color video frames at 60 Hz<br>**Action:** discrete, variable for each game<br>**Reward:** rescaled game scores<br>**RL approach:** deep Q-learning |
| **(B) Flight control:** Stratospheric Balloons ||
| Stratospheric balloons are high-altitude balloons that can reside 15 to 60 kilometers above sea level for months at a time. These balloons carry up to 1.1 tons of payload, typically tools used for weather forecasting, satellite navigation, atmospheric chemistry experiments, and testing new space technology. Navigation in these high-altitude settings is dependent on stratospheric winds, of which relevant meteorological data is sparse, and solar availability, which is needed to charge the battery. Model-free reinforcement learning enabled effective navigation over the Pacific Ocean over 39 days, using distributed Q-learning (46). | **Key challenges addressed:** incomplete data, noisy observations, unreliable solar availability, safe navigation, long planning horizon<br>**Objective:** navigate a super-pressure balloon to float near weather station<br>**State:** 1,083 wind variables. 16 ambient variables<br>**Action:** discrete (Ascend, descend, stay)<br>**Reward:** distance from a weather station, with maximum reward within 50 km of the station<br>**RL approach:** Model-free Q-learning with experience replay |
| **(C) Clinical Decision-Making:** Sepsis Treatment ||
| Sepsis is a life-threatening excessive immune response to an infection that may lead to organ failure or death. Treating sepsis involves a complex mix of antibiotics, corticosteroids, timing and dosage of drugs, and intravenous fluids. The treatment regiment for patients in the intensive care unit (ICU) must be customized to each individual patient in response to that patient's response to medical interventions. Deep Q-learning helped learn treatment policies to help reduce patient mortality by 1.8–3.7% (47). | **Key challenges addressed:** continuous state, sparse reward signals, stochasticity, delayed feedback, interpretability<br>**Objective:** improve patient survival<br>**State:** continuous. 48 values of demographics, physiological data, and vital signs)<br>**Action:** discrete. 5×5 intervention options with different amounts of IV fluid and vasopressor dosage |

| | **Reward:** weighted sum of indicators of patient health including the extent of organ failure and changes in blood pressure <br> **RL approach**: dueling double-deep deep-Q learning |
|---|---|

# Possibilities and pitfalls of applying RL to adaptive management problems

If real-world complexity has forced the practice of adaptive management beyond the reach of theory, emerging paradigms of reinforcement learning appear to at last be putting such challenges within reach of algorithms (Table 1). But just because this may be possible, is it really a good idea? Adapting reinforcement learning for adaptive management could open possibilities (20,40) but also introduces new pitfalls while re-surfacing age-old concerns of algorithmic decision processes (48). We divide these possibilities and perils into three themes. First, the conceptual shift of an RL approach to **learning** as something based on heuristics and experience rather than rigorous mathematical theorems. This can recapitulate some of the benefits but also the shortcomings of human decision-making. Second, RL still faces the same challenge of any objective-based decision-making: accurately defining the task at hand. RL overcomes some computational constraints but still requires defining a scope of possible rewards, states, and actions. We refer to this as **world-making**. Third, deep RL's technical and computational needs may limit its application to the largest technology institutions with access to these resources. This **stakeholder shift** exacerbates potential ethical and political consequences. Here we outline both the technical and social components of these opportunities and challenges across the three main themes of (1) learning, (2) worldmaking, and (3) shifting stakeholders (Figure 2). We hope both the RL and adaptive management communities recognize and focus on addressing these challenges when developing and implementing environmental decision-making.

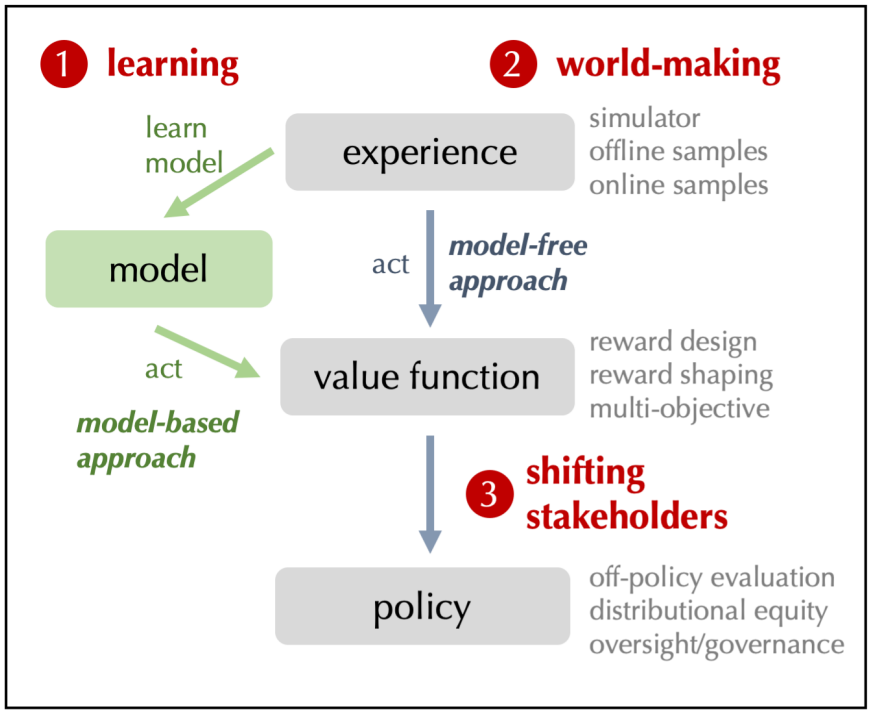

**Figure 2:** Traditional adaptive management relies on modeling the environment using Markov decision processes, which mirrors a "model-based" approach to RL (green pathway, left). Model-free RL (grey pathway, center) eschews learning an intermediate model to instead directly estimate the reward for taking specific actions at a given state. As we outline, reinforcement learning brings both promise and new challenges for adaptive management for learning, world-making, and shifting stakeholders, which all impact different components of the RL pipeline.

**Table 1:** Emerging methods and approaches in RL have the potential to help overcome a myriad of challenges in adaptive management, from making effective decisions in complex environments over long planning horizons to learning strategies in high-stakes situations. The following methods help define a research agenda for RL researchers seeking to contribute to environmental management.

| Adaptive management challenge | RL methods and concepts | Citation |
|---|---|---|
| *Dealing with uncertainty* | | |
| Data used to train RL agents may not be precise or generalize to real-world settings. | *robust RL* aims to learn policies that perform well across a large class of possible environments, including environments that may not have been explicitly encountered during training. | (49–51) |
| When data is insufficient to reliably predict the outcome of an action, decision makers wish to understand the degree of uncertainty. | *uncertainty quantification* measures and attempts to reduce uncertainty in predictive systems. These uncertainty estimates can be used to constrain the RL policy to avoid taking actions with high uncertainty in outcome, related to safe RL. | (52–54) |
| A given model does not precisely describe the true dynamics of the ecological system, regardless of what values are used to instantiate the parameters. | *model misspecification* refers to RL settings in which the Markov decision process used to model the environment does not describe reality. Model misspecification is typically addressed with robust RL, including model-free approaches. | (55–57) |
| For a given set of observed data, multiple sets of parameters used to instantiate the model may have the same probability distribution of being the best-fit model. | *non-identifiability* describes the challenge of learning and planning under in the presence of unmeasured state variables (confounders). | (58,59) |
| The reward function may be unknown in advance. | *inverse RL* attempts to recover the reward function given an optimal policy and environment dynamics. | (60,61) |
| *Limited opportunities to interact with real-world, high-stakes settings* | | |
| Must learn optimal policies using historical data, without collecting new data. This challenge arises when data collection is expensive — as is often the case in conservation management. | *offline RL* (or *batch RL*) learns the best policy possible given historical observations (a static dataset) without exploration. Requires *off-policy evaluation* to estimate the performance of policies that were never enacted in the historical data. | (41) |
| In high-stakes settings, managers wish to be risk-averse to avoid potentially catastrophic settings, such as unintentionally wiping the population of one species. | *RL safety* trains an RL policy by limiting explorative actions to those that are unlikely to reach very bad states, for example by imposing additional constraints or avoiding actions with high uncertainty. | (62,63) |
| Decision makers wish to know why a policy calls for a specific action. | *explainability* in RL aims to provide human-understandable explanations for why an RL method recommends a specific action, in contrast to "black box" methods. | (64,65) |
| Decision makers want to have human experts oversee and possibly override an RL agent's decisions. | *human-in-the-loop* systems treat humans as experts and defer to these human experts to make decisions when the RL agent is highly uncertain. | (66,67) |
| *Complex environments with long planning horizons* | | |
| The underlying rules of the environment may be changing over time, due to exogenous factors such as climate change or socioeconomic drivers. | *non-stationarity* refers to changes over time in the underlying dynamics of the MDP. In dynamic settings, the state constantly changes, but with non-stationarity the transition dynamics and rewards shift as well. | (68,69) |
| Management decisions may be highly complex, thus difficult to learn from scratch. | *curriculum learning* trains an agent on progressively harder tasks, using *transfer learning* to build off knowledge learned from previous tasks to subsequent tasks. | (34,70) |
| Management decisions may require planning challenging multi-step tasks over long time horizons. | *hierarchical RL* decomposes long-horizon tasks into more tractable subtasks. | (71) |
| After we receive a reward, we want to know which action(s) were critical | *credit assignment* evaluates the utility of individual actions over a long | (72–74) |

| to that outcome. This challenge is exasperated by *delayed feedback*. | sequence of steps. | |
| --- | --- | --- |
| Rewards may be sparse, especially with *delayed feedback*, and the benefit of intermediate actions may not be immediately obvious. | *reward shaping* provides more gradual, localized feedback to guide the policy toward high-reward states. | (75–77) |
| *Multiple stakeholders, multiple agents* | | |
| Multiple stakeholders may each have their own objective. | *multi-objective RL* learns optimal decisions in the face of multiple conflicting objectives. This challenge is most salient when the relative weighting (importance) of the objectives is not known. | (78,79) |
| Multiple agents may be acting in an environment simultaneously. | *multi-agent RL* trains agents to act in the presence of other agents. In a *cooperative* setting, these agents share a goal but in a *competitive* setting, these agents have non-aligned goals which may be in conflict. | (80,81) |

## Learning

The central difference between model-free deep reinforcement learning and the theory of adaptive management regards *learning*. In adaptive management, learning is defined as reducing uncertainty over parameters or candidate predictive models[1] of the underlying ecological processes (Figure 1A). Learning is expressed in terms of quantitatively precise probability distributions and realized through mathematically precise theorems such as Bayes' rule to dictate how model 'beliefs' (probability distributions) narrow in response to actions and new information (Figure 1A). By contrast, a human manager does not necessarily need a predictive model of the process to adjust a decision (Figure 1B). It is possible to propose a policy without a model based on experience alone. For example, if the estimated waterfowl population (Box 1B) decreased too much last year compared to the year before, it's probably a good idea to lower the harvest quota this year. Of course, the theory may give the same answer with more quantitative precision – how much to lower the quota (and also just how much to re-adjust "belief" probabilities towards some more pessimistic growth rates of the species) – but that answer is only as good as the models it considers. The manager's experience may factor in variables ignored by the models – a harvest quota of zero may be socially or politically unacceptable, while past experience of ups and downs may provide an experienced manager with a notion for the right size of adjustment with nary an equation (82,83).

Model-free RL capitalizes on this experience-driven approach to decision-making. The RL agent does not need to predict future states; it decides what action to take given only experience from past states and resulting rewards (or costs). The RL researcher places an untrained agent in a novel environment (usually a computer simulation, e.g., in the Atari example (Box 2A)), in which the agent takes exploratory actions while adjusting its policy to improve long-term reward. From repeated simulations over hundreds of thousands of episodes, the agent will extensively explore the space of actions and outcomes. The process is sometimes compared to a newborn first exploring the world around them. Like the newborn, this RL agent is not entirely naive – the researcher must select among a myriad of specific algorithms each with very different approaches to solving the reinforcement learning problem. The researcher, just like the parent of a newborn, may present a modified system of rewards and costs to coax along the desired learning more efficiently in a process called *reward shaping* (Table 1). Reward shaping becomes particularly useful when there is a big payoff only after a long sequence of actions (e.g., rewarding distance to the end of a maze rather than only completion).

---

[1] Some authors distinguish between model uncertainty that is 'structural' in nature, e.g., if recruitment follows a Ricker-shaped curve or a Beverton-Holt shaped curve, versus uncertainty that is only of a 'parametric' nature – e.g., the value of initial growth rate "r" in a Ricker model. In practice the lines are blurry as it is often possible for a structurally flexible enough model to represent both families of curves in terms of the choice of some additional parameters. In fact, the deep neural networks underlying most modern machine learning including RL-based methods owe their success to being precisely such highly flexible function approximators. The key observation of model-free RL is that the functions seek to approximate are not the process itself – the probability from any possible current state to any possible future state under any possible action – but rather, the often smaller map between possible states to the space of possible actions – the 'policy function' or 'value function' the manager should adopt.

When trained in a single environment, the strategies that RL agents learn rarely generalize to even small deviations from past experience. The agent will often *overfit* to the smallest details – the units of measurement, the duration of the particular episode. More generally applicable strategies can be found by presenting the agent with a wide variety of environments. For example, one pervasive challenge in learning from simulated experience is the "sim2real" gap, the difference between an RL agent's performance in a simulated environment versus a real environment (84). Robust RL techniques may help close the sim2real gap and avoid overfitting (85) (Table 1). A wealth of emerging approaches seek to improve generalizability. *Curriculum learning* algorithms seek to provide the most efficient way to interleave different environments (Table 1). In *adversarial learning*, a second agent seeks to learn and propose alterations to the environment that are most likely to fool the focal agent into poor performance.

Most large-scale ecological simulation systems still fall short of capturing the many processes involved (40) but are already beyond the reach of dynamic programming methods of classical adaptive management. Collectively, advances in the ecological realism of simulations and computational RL methods make it feasible to train intelligent agents across a wide variety of simulated environments. Historical observations can only be paired with historical actions, and thus never provide an agent much insight into the outcomes of the actions not taken but can nevertheless be used to supplement and ground-truth training based on simulation (Table 1). A more fraught question concerns the role of RL-based learning in real-world contexts. Like the distinction between active and passive adaptive management, RL is typically divided between 'training' and 'evaluation' loops. In the training loop, the RL agent explores their action space to discover and adjust their decision strategy ('active learning'). In the evaluation loop, acting is essentially passive, with the RL agent seeking to maximize expected utility without updating their decision strategy. Evaluation need not always be passive in RL (especially in 'low-stakes' real-world scenarios, such as a physical robot learning to walk or handle objects) but mirrors the general preference of managers to rely on passive adaptive management in high-stakes scenarios.

## World-making

While RL might allow quantitative adaptive management to consider more realistic state and action spaces, reducing the numerical constraints on problems only refracts the issue of distilling a complex environment: how do we bound an environmental state, define management objectives, and determine a set of available actions while ensuring these represent environmental realities and values of those most impacted by the decisions? How do we create sufficient simulations or decide on the appropriate data streams to train algorithms with? RL might expand the scope and range of problems that we can solve, but it does not remove the sociopolitical considerations inherent to how those problems are defined.

Let's imagine reformulating the waterfowl harvest problem as a RL problem. We could simply create a simulation of the states and their responses to actions in alignment with the current model-based formalization (Box 1B): the action space is an annual harvest quota to maximize expected long-term yields and the state space is a one-dimensional representation of waterfowl stock. But given fewer constraints, we might represent the waterfowl stock as part of a larger ecological (or socio-ecological) system responding to human land use, climatic shifts, and weather extremes, overcoming shortcomings of current methods, like the consideration of nonautonomous systems. Even if the action space remained the same (harvest quota), the algorithm's optimal policy would change. The action space could similarly be expanded to better capture the possibility of decisions (e.g., to a temporally and spatially dynamic closure rather than a single harvest quota), changing not only the policy and its impact on resource users, but the underlying system trajectory. These changes seem benign, if not beneficial, but it is easy to envision how the imagination and values of the algorithm designer impact not only the conceptualization of the environment (state spaces), but the solutions derived, and actions taken, which ultimately feedback to create the reality of that system (48). Leaving us to question what parts of the system are included in the simulations and how that might shift the distribution of benefits and costs.

Barriers to the adoption of adaptive management strategies arise not only from a lack of realism in system formalization or capacity to deal with complexity, but also from disagreement over whose values are represented in the decision objectives and the potential risks of following algorithmic suggestions (86). RL does not sidestep these issues, but methods such as *multiobjective RL* (Table 1) can learn optimal decisions in the face of multiple conflicting objectives, and *inverse reinforcement learning* can help align the values as they are represented in formalized rewards with real-world values ("value alignment") (60,61) (Table 1). Additionally, *reward shaping* (Table 1) can help to ensure RL agents do not myopically take actions that lead to short-term gain over long-term benefits.

Even in light of technical methods to improve the alignment of values and capture multiple objectives, the issues of world-making are social and normative at their core. Specifying state, action, and rewards in RL applications will necessarily reflect both epistemic values and contextual values of the developer (87,88). Which begs the question: *who has the power and capacity to define problems and develop RL algorithms?*

## Shifting stakeholders

Given both the technical expertise and computational requirements needed to train RL algorithms, industry (specifically big tech) involvement in the development and deployment of these methods is commonplace across environmental application domains (89). The shift from government-maintained and managed algorithms - as is currently the case in most environmental adaptive management contexts (such as waterfowl adaptive management; Box 1B) - to industry-maintained algorithms would create a new set of actors in the environmental regulatory processes. Because political and financial concerns may influence the design of RL environments and agents, developing transparent and inclusive participatory processes will be critical to ethical and equitable development and application of RL to adaptive management problems.

Beyond shifting power and creating new environmental actors, RL-derived environmental decisions risk undermining trust in environmental governance systems by increasing the ambiguity of who is accountable for future environmental degradation (87). If a decision is derived from an algorithm that relies on trial and error rather than clearly mapping to a model choice, are poor outcomes anyone's fault? Moreover, if an RL agent continues to learn and adapt while interacting with the system (adaptive management, by definition), how do we ensure that its policies are meaningfully overseen (88)? In this way, RL differs from more transparent model-based methods in the relative lack of capacity to query solutions, potentially obscuring biases and compliance with regulations (90). Lessons from RL applications to other safety-critical domains, such as nuclear fusion (91), and tools from the explainable AI subfield (92,93), might help mitigate these issues (Table 1). However, problems of explainability and safety become even more pronounced when RL is proposed for controlling less identifiable and high-dimensional systems, as is the case in many environmental management contexts.

Ethical AI principles provide some guidance to procedure and practice to ensure safe application of algorithms. But these guidelines, like the algorithms themselves, are primarily developed in the Global North, notably missing perspectives from Central and South America, Africa, and Central Asia (94). Moreover, ethical guidelines rarely address the many dimensions of power implicated in world making; not only the power to make decisions or define objectives, but power to set the agendas (e.g., defining objectives, state and action spaces) and shift ideologies (48). Applying decolonial theories to AI application and development, as discussed in (95), might help address the shortcomings of AI ethics and recenter the importance of power and representation in procedural and development processes.

While the technical synergies and differences between reinforcement learning and model-based adaptive management methods are outlined throughout this paper, simultaneously considering the parallels between AI ethics (94), science technology studies (96), and political ecology (97) are critical when considering applications of RL to safety critical real-world domains like environmental management.

# Conclusion

To bridge the gap between the science and practice of adaptive management there is a need for decision-centered methods that capture the complexity and uncertainty of ecosystems. Advances in deep learning have positioned *reinforcement learning* (RL) as a promising approach to solve sequential problems under uncertainty, while sidestepping the need to define a set of candidate models or effectively refine our belief in those models. Here we highlight recent advances in RL methods that overcome several limitations — such as high-dimensional spaces, imperfect models, and lack of accurate simulators — that have prevented adaptive management from moving beyond theory in complex situations. Simultaneously we underline key priorities for RL — such as robustness, safety, and multi-objective rewards — to enable its effective and responsible deployment for ecological decision-making.

# Funding

This material is based upon work supported by the National Science Foundation under Grant No. DBI-1942280. Xu was supported by a Google PhD Fellowship.

26. Irwin, Elise R., et al. "Adaptive management of flows from RL Harris dam (Tallapoosa river, Alabama)-stakeholder process and use of biological monitoring data for decision making." *Open-File Report-US Geological Survey* 2019-1026 (2019).
27. McGOWAN, CONOR P., et al. "Multispecies modeling for adaptive management of horseshoe crabs and red knots in the Delaware Bay." *Natural Resource Modeling* 24.1 (2011): 117-156.
28. Johnson, Fred A., et al. "Multilevel learning in the adaptive management of waterfowl harvests: 20 years and counting." *Wildlife Society Bulletin* 39.1 (2015): 9-19.
29. Ogden, Aynslie E., and John L. Innes. "Application of structured decision making to an assessment of climate change vulnerabilities and adaptation options for sustainable forest management." *Ecology and Society* 14.1 (2009).
30. Kingsford, Richard T., Harry C. Biggs, and Sharon R. Pollard. "Strategic adaptive management in freshwater protected areas and their rivers." *Biological Conservation* 144.4 (2011): 1194-1203.
31. Agrawal, Arun. "Adaptive management in transboundary protected areas: The Bialowieza National Park and Biosphere Reserve as a case study." *Environmental Conservation* 27.4 (2000): 326-333.
32. Fonnesbeck, Christopher J. "Solving dynamic wildlife resource optimization problems using reinforcement learning." *Natural Resource Modeling* 18.1 (2005): 1-40.
33. Chadès, Iadine, et al. "Optimization methods to solve adaptive management problems." *Theoretical Ecology* 10.1 (2017): 1-20.
34. Silver, David, et al. "Mastering the game of Go with deep neural networks and tree search." *nature* 529.7587 (2016): 484-489.
35. Won, Dong-Ok, Klaus-Robert Müller, and Seong-Whan Lee. "An adaptive deep reinforcement learning framework enables curling robots with human-like performance in real-world conditions." *Science Robotics* 5.46 (2020): eabb9764.
36. Yu, Chao, et al. "Reinforcement learning in healthcare: A survey." *ACM Computing Surveys (CSUR)* 55.1 (2021): 1-36.
37. Perera, A. T. D., and Parameswaran Kamalaruban. "Applications of reinforcement learning in energy systems." *Renewable and Sustainable Energy Reviews* 137 (2021): 110618.
38. Neftci, Emre O., and Bruno B. Averbeck. "Reinforcement learning in artificial and biological systems." *Nature Machine Intelligence* 1.3 (2019): 133-143.
39. Zheng, Stephan, et al. "The AI Economist: Taxation policy design via two-level deep multiagent reinforcement learning." *Science advances* 8.18 (2022): eabk2607.
40. Urban, Mark C., et al. "Coding for Life: Designing a Platform for Projecting and Protecting Global Biodiversity." *BioScience* 72.1 (2022): 91-104.
41. Levine, Sergey, et al. "Offline reinforcement learning: Tutorial, review, and perspectives on open problems." *arXiv preprint arXiv:2005.01643* (2020).
42. Kostrikov, Ilya, Ashvin Nair, and Sergey Levine. "Offline reinforcement learning with implicit q-learning." *arXiv preprint arXiv:2110.06169* (2021).
43. Oliver, Ruth, et al. "Camera trapping expands the view into global biodiversity and its change" (2023)
44. Jiang, Nan, and Lihong Li. "Doubly robust off-policy value evaluation for reinforcement learning." *International Conference on Machine Learning*. PMLR, 2016.
45. Gottesman, Omer, et al. "Evaluating reinforcement learning algorithms in observational health settings." *arXiv preprint arXiv:1805.12298* (2018).
46. Bellemare, Marc G., et al. "Autonomous navigation of stratospheric balloons using reinforcement learning." *Nature* 588.7836 (2020): 77-82.
47. Raghu, Aniruddh, et al. "Deep reinforcement learning for sepsis treatment." *arXiv preprint arXiv:1711.09602* (2017).
48. Scoville, Caleb, et al. "Algorithmic conservation in a changing climate." *Current Opinion in Environmental Sustainability* 51 (2021): 30-35.
17

74. Harutyunyan, Anna, et al. "Hindsight credit assignment." *Advances in neural information processing systems* 32 (2019).
75. Laud, Adam Daniel. *Theory and application of reward shaping in reinforcement learning*. University of Illinois at Urbana-Champaign, 2004.
76. Hu, Yujing, et al. "Learning to utilize shaping rewards: A new approach of reward shaping." *Advances in Neural Information Processing Systems* 33 (2020): 15931-15941.
77. Marom, Ofir, and Benjamin Rosman. "Belief reward shaping in reinforcement learning." *Proceedings of the AAAI Conference on Artificial Intelligence*. Vol. 32. No. 1. 2018.
78. Liu, Chunming, Xin Xu, and Dewen Hu. "Multiobjective reinforcement learning: A comprehensive overview." *IEEE Transactions on Systems, Man, and Cybernetics: Systems* 45.3 (2014): 385-398.
79. Mossalam, Hossam, et al. "Multi-objective deep reinforcement learning." *arXiv preprint arXiv:1610.02707* (2016).
80. Shoham, Yoav, Rob Powers, and Trond Grenager. *Multi-agent reinforcement learning: a critical survey*. Technical report, Stanford University, 2003.
81. Zhang, Kaiqing, Zhuoran Yang, and Tamer Başar. "Multi-agent reinforcement learning: A selective overview of theories and algorithms." *Handbook of Reinforcement Learning and Control* (2021): 321-384.
82. McDonald-Madden, E. V. E., Peter WJ Baxter, and Hugh P. Possingham. "Subpopulation triage: how to allocate conservation effort among populations." *Conservation Biology* 22.3 (2008): 656-665.
83. Chadès, Iadine, et al. "General rules for managing and surveying networks of pests, diseases, and endangered species." *Proceedings of the National Academy of Sciences* 108.20 (2011): 8323-8328.
84. Höfer, Sebastian, et al. "Sim2Real in robotics and automation: Applications and challenges." *IEEE transactions on automation science and engineering* 18.2 (2021): 398-400.
85. Thomas, Philip, and Emma Brunskill. "Data-efficient off-policy policy evaluation for reinforcement learning." *International Conference on Machine Learning*. PMLR, 2016.
86. Allen, Craig R., and Lance H. Gunderson. "Pathology and failure in the design and implementation of adaptive management." *Journal of environmental management* 92.5 (2011): 1379-1384.
87. Scoville, Caleb, et al. "Algorithmic conservation in a changing climate." *Current Opinion in Environmental Sustainability* 51 (2021): 30-35.
88. Gilbert, Thomas Krendl, et al. "Reward Reports for Reinforcement Learning." *arXiv preprint arXiv:2204.10817* (2022).
89. Verdegem, Pieter. "Dismantling AI capitalism: the commons as an alternative to the power concentration of Big Tech." *AI & society* (2022): 1-11.
90. Puiutta, Erika, and Eric Veith. "Explainable reinforcement learning: A survey." *International cross-domain conference for machine learning and knowledge extraction*. Springer, Cham, 2020.
91. Degrave, Jonas, et al. "Magnetic control of tokamak plasmas through deep reinforcement learning." *Nature* 602.7897 (2022): 414-419.
92. Roscher, Ribana, et al. "Explainable machine learning for scientific insights and discoveries." *Ieee Access* 8 (2020): 42200-42216.
93. Bhatt, Umang, et al. "Explainable machine learning in deployment." *Proceedings of the 2020 conference on fairness, accountability, and transparency*. 2020.
94. Jobin, Anna, Marcello Ienca, and Effy Vayena. "The global landscape of AI ethics guidelines." *Nature Machine Intelligence* 1.9 (2019): 389-399.
95. Mohamed, Shakir, Marie-Therese Png, and William Isaac. "Decolonial AI: Decolonial theory as sociotechnical foresight in artificial intelligence." *Philosophy & Technology* 33.4 (2020): 659-684.
19

96. Bareis, Jascha, and Christian Katzenbach. "Talking AI into being: The narratives and imaginaries of national AI strategies and their performative politics." *Science, Technology, & Human Values* 47.5 (2022): 855-881.
97. Nost, Eric, and Emma Colven. "Earth for AI: A Political Ecology of Data-Driven Climate Initiatives." *Geoforum* 130 (2022): 23-34.